\documentclass{desyproc}

\usepackage{hyperref}

\begin{document}

\title{\vspace{-3cm}
\hfill{\small{DESY 08-192}}\\[2cm]
Status of the ALPS Experiment}

\author{{\slshape Klaus Ehret$^1$ for the ALPS Collaboration (http://alps.desy.de/)}\\[1ex]
$^1$DESY, Notketra{\ss}e 85, 22607 Hamburg, Germany }

\contribID{ehret\_klaus}

\desyproc{DESY-PROC-2008-02}
\acronym{Patras 2008} 
\doi  

\maketitle

\begin{abstract}
The ALPS experiment at DESY searches for light particles which are 
coupling very weakly to photons. 
Primary physics goal is the search for axion like particles 
in a photon regeneration experiment. 
Central part of the experimental setup is a five Tesla 
strong superconducting HERA dipole magnet. 
During two operation periods in the years 2007 and 2008 
we have collected first data and explored the sensitivity of the setup. 
A Fabry Perot laser cavity is being set up in order to increase the 
sensitivity by more than one order of magnitude.
\end{abstract}

\section{Physics Motivation and Goals of ALPS}
One of the most exciting quest in particle physics is the search for new particles beyond the standard model.
Extensions of the standard model predict not only new particles with masses above the electroweak scale ($\sim 100$ GeV) but also the so called WISPs ({\bf{W}}eakly {\bf{I}}nteracting {\bf{S}}ub-eV {\bf{P}}articles). 
Several possible candidates are axions~\cite{Weinberg:1977ma} or axion like particles (ALPs), light spin 1 particles called ``hidden sector photons"~\cite{Okun:1982xi} or light minicharged particles~\cite{Holdom:1985ag}. For a summary on their role in physics beyond the standard model see the contributions of J.~Jaeckel and A.~Ringwald to these proceedings~\cite{Ringwald:2008cu}.
It is certainly an important and fundamental question whether any of these light particles exist.
Unfortunately, the predictions for the masses and couplings of WISPs are typically not precise and extensive searches in broad parameter spaces have to be performed.
Of course, any experimental measurement which gives new indications or new limitations is highly welcome. 
Nowadays, the strongest constraints come from astrophysical and cosmological arguments (see the contribution of J.~Redondo in these proceedings~\cite{Redondo:2008en}) and from dedicated laboratory experiments~\cite{Amsler:2008zz}.
Very recently, the observations of PVLAS~\cite{Zavattini:2005tm} triggered the interest, exploration and setup of new  low energy experiments using high photon fluxes combined with strong electromagnetic fields.

The {\bf ALPS} experiment, located at DESY in Hamburg, uses a spare superconducting HERA dipole magnet and a strong laser beam for ``{\bf{A}}xion {\bf{L}}ike {\bf{P}}article {\bf{S}}earch". In fact, the experiment has also a large sensitivity for other WISPs so the acronym ALPS should stand more precisely for ``{\bf{A}}ny {\bf{L}}ight {\bf{P}}article {\bf{S}}earch".
The primary goal is the indirect detection of a light ALP in a ``light shining through a wall" (LSW) experiment~\cite{Sikivie:1983ip}. 
A small fraction of the incident photons from our laser can convert to ALPs $\phi$ in the presence of the magnetic field by the so-called Primakoff effect~\cite{Primakoff:1951ww}. 
Being $\phi$s very weakly interacting with ordinary matter, they cross light-tight walls without significant absorption. 
Behind the absorber, some of these ALPs will reconvert via the inverse-Primakoff process into photons with the initial properties.

The probability of the Primakoff transition $P_{\gamma \to \phi}$ is the same than for the inverse-Primakoff $P_{\phi\to \gamma}$ in the ALPs symmetric setup. 
Therefore the LSW probability is just the square of $P_{\gamma \rightarrow \phi} = g^{2} B^2 E^2/(4 m_\phi^4) \sin^2 m_\phi^2 L/(4 E)$ with $B$ the magnetic field strength, $L$ the length of the conversion region and $E$ the photon energy. 
The ALP parameters: mass ($m_\phi$) and two photon coupling ($g$) are assumed to be unrelated (see~\cite{Ringwald:2008cu} for the relevant definitions).

The transition probability is maximal in the limit $m^2_\phi/E \to 0$ where it is coherent along the full length, reaching $g^2B^2L^2/4$. The mass reach of the experiment is determined therefore by $E$ through the coherence condition $m^2_\phi < 2 \pi E / L $. 
The polarization of the beam allows to distinguish between scalar and pseudo scalar ALPs.

After the submission of the Letter of Intent~\cite{Ehret:2007cm} the ALPS experiment was approved by the DESY directorate in January 2007. The ALPS collaboration comprises also the Laser Zentrum Hannover (LZH), the Hamburg observatory (HO) and the Albert Einstein Institute(AEI).

\section{The Experimental Setup}
\begin{figure}[t]
\centerline{\includegraphics[width=1.0\textwidth]{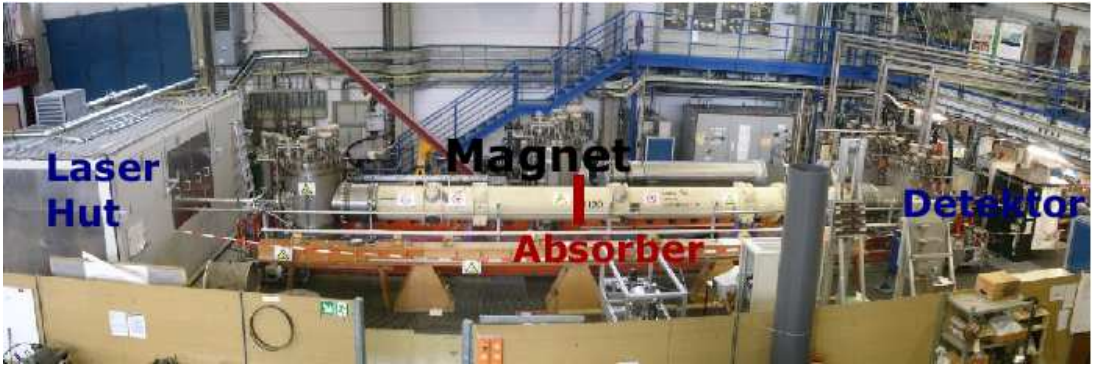}}
\caption{Photo of the ALPS Experiment, showing the laser hut, the superconducting magnet and the cabinet housing the detector setup.}\label{Fig:PhotoSetup}
\end{figure}

\begin{figure}[t]
\vspace{-0.5cm}
\centerline{\includegraphics[width=0.55\textwidth]{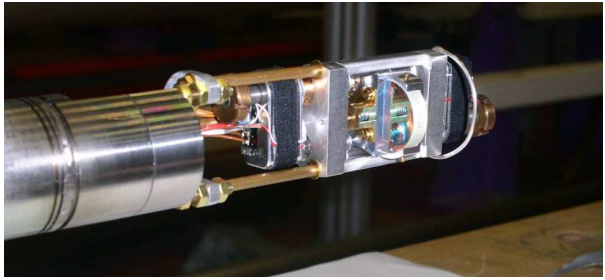}}
\caption{Adjustable mirror with custom made pico-motors attached to the laser tube.}\label{Fig:Setup}
\end{figure}

The ALPS experiment is built up around the HERA dipole which features a field $B=5.16$~T and a length of 8.42~m, cf.~~Fig.~\ref{Fig:PhotoSetup}. 
Its beam pipe is bent with a remaining clear aperture of only 18~mm, which implies serious demands on the beam quality of the laser. The interior is insulated against the cold part of the magnet, allowing to perform the experiment at room temperature. 
In order to keep the temperature stable the beam pipe is flushed with nitrogen.
Inside the dipole beam pipe we place two further tubes which bound the $\gamma-\phi$ conversion and reconversion regions and are operated under vacuum conditions. This is crucial since an index of refraction $n>1$ suppresses the conversion probabilities~\cite{Raffelt:1987im}.
Both tubes range from either side approximately to the middle of the magnet and can be easily removed.
A removable light-tight absorber wall is mounted on the inner end of the {\bf detector tube} while an adjustable mirror is attached to the inner side of the {\bf laser tube}, cf. Fig.~\ref{Fig:Setup}, because dissipating the high laser power inside the magnet could produce dangerous quenches of the magnet. On both sides of the laser tube there are vacuum sealed windows. 
Custom made pico-motors based on piezo actuators allow a very precise remote adjustment of the inner mirror within the strong magnetic field. This was a very important precondition for the later setup of a laser cavity inside the superconducting magnet.

\subsection{The Laser}
The laser setup, built on an solid optical table inside our laser hut, allows to adjust the intensity and the polarization of the beam. A low intensity reference beam is guided in a beam tube outside the magnet parallel to the main beam to the detector setup.
Due to the small aperture of the magnet and the requirement to focus the beam on a few pixels of the used camera a small laser beam width with a very good beam quality factor $M^2\sim 1$ is required. 
Both, the beam divergence $\theta$ and the minimal spot size $\sigma_{min} $ are proportional to $M^2$.

There exist high power infrared lasers but the demanding request on the beam quality excludes many industry devices. 
Furthermore, it is a challenging enterprise to get affordable high efficient and low noise detectors for infrared photons. 
In summer 2007 we setup the experiment with a 3.5 W Verdi-Coherent CW green laser with $\lambda = 532$~nm and  $M^2<1.1$ and a commercial camera with good efficiency for green light. Based on the experience of a test run in autumn 2007 we abandoned the initial plan to use an infrared laser. 

By the end of the year 2007 we setup and operated successfully a LIGO type pulsed laser system~\cite{Fredeetal} with a pulse-length of 15~nsec and a repetition rate of 20~kHz. The system delivered 14~W of green laser light,
which corresponds to a photon flux of $4\cdot10^{19}s^{-1}$. During an extensive commissioning phase in spring 2008 we operated the complete setup and explored its sensitivity (details are discussed later on).
Based on this experience, it became clear that we had to increase the photon flux by one order of magnitude to become competitive with other experiments and to reach a sensitivity which allows the exploration of yet untouched areas in the parameter phase space of ALPs (cf.~Fig.~\ref{Fig:Sens}).
Together with our new collaborators from the AEI we setup a resonant optical cavity in the first half of the HERA magnet. 
An enhanced LIGO (eLIGO) laser, developed at the LZH for gravitational experiments delivers 35~W 1064~nm laser light with excellent beam characteristics. 
This is converted into 532~nm green laser light and fed into the {8.62}~m long cavity, which is bounded by the outer mirror, mounted on the optical table inside the laser hut, and the mirror at the inner end of the laser beam tube cf.~Fig.~\ref{Fig:Setup}. 

The resonant cavity is locked by adopting the frequency of the eLIGO laser to compensate the length fluctuations of the setup.
It was a major success and very important proof of principle, that the cavity could be locked over days and operates stable also when the magnet is powered. 
Recently we studied details of this ambitious setup in order to improve the performance and to exploit the full capability.
Two major upgrade steps are under preparation: a second resonant laser cavity around the frequency doubling crystal 
and the inclusion of the outer mirror into the cavity vacuum system.
You may find several details concerning the laser setup and cavity in the contributions of
M.~Hildebrand to these proceedings\cite{PatrasWSDESY}.

\subsection{Detector}
As photon detector we used so far the commercial astronomy CCD camera SBIG ST-402ME with $765 \ast 510$ $9~\mu$m~$\times ~9~\mu$m pixels and a quantum efficiency of $60\%$ for $\lambda = 532$~nm. 
We operate the camera at $-5 °$~C. It has a low dark current and a small readout noise of $17 ~e^-$. The camera allows sampling times between 0.04~s and 1~h. 
The beam is focused on a small area $\approx 10~ \mu\rm{m}$, i.e. a few pixels.

In order to improve the sensitivity we ordered the camera PIXIS 1024-BL (Princeton Instruments) with a quantum efficiency of $95~ \%$ for $\lambda = 532$~m. This camera operates at $-70 °$~C, featuring a lower dark current of $0.001~e^-$ pixel$^{-1}s^{-1}$ and a readout noise of less than $4~e^-$.
The camera was delivered by the end of August and its performance is now under investigation.

\section{Commissioning Run 2008 \& Exploration of Sensitivity}
During the commissioning run in spring 2008 we used the 14~W LIGO laser and the SBIG camera to collect around $100$~h of data with magnet and laser on.
The magnet, the  camera and the laser worked very reliably.
In order to minimize the impact of readout noise we used a sampling time of 20~min or one hour. 
The absorber was removed a few times in order to test the alignment.
Unfortunately the front and back surfaces of the mirror inside the magnet were not sufficiently parallel, causing a deflection of the light passing the mirror used for the alignment of the setup.
This limits the knowledge of the beam spot, i.e.~the position of potential re-converted photons on the CCD and prohibits to use the data for real physics. 
The troubling mirror was immediately replaced.

The acquired data were nevertheless used to go through the complete data analysis chain.
Around 10~\% of the frames are rejected by visual inspection because they contain intense tracks (likely from ambient radioactivity or cosmics) close to the signal region. 
As measurement we use the sum of the pixel values in a $3\times 3$ array around the beam spot region.
After the correction of baseline shifts, which are presumably correlated with varying temperatures of the surroundings, the remaining fluctuations corresponds to the expectations of uncorrelated dark currents and the read-out noise of the individual pixels.
The classification of data depends on the physics. Data with magnet on and vertical (horizontal) laser polarization are signals for pseudo-scalar (scalar) ALPs. All other data including dark frames are background.
The final observable is the difference in the mean of the distribution of many signal measurements to the mean 
of the distribution of many background measurements. 
We see no signal, i.e.~no significant difference between signal and background distributions. 

Taking into account the conversion factors and a conservative estimate for the efficiencies 
the sensitivity for the reconverted photon flux is around  $40$~mHz. This leads together with the initial laser photon flux
of $4\cdot10^{19}\gamma/$s to a detection probability $P_{\gamma \rightarrow \phi  \rightarrow \gamma} \approx 10^{-21}$. 
The deduced sensitivity in the ALP and in the ``hidden sector photon" parameter space is plotted in Fig.~\ref{Fig:Sens}.

\section{Outlook}
\begin{figure}[t]
\vspace{-0.5cm}
\centerline{
\includegraphics[width=0.5\textwidth]{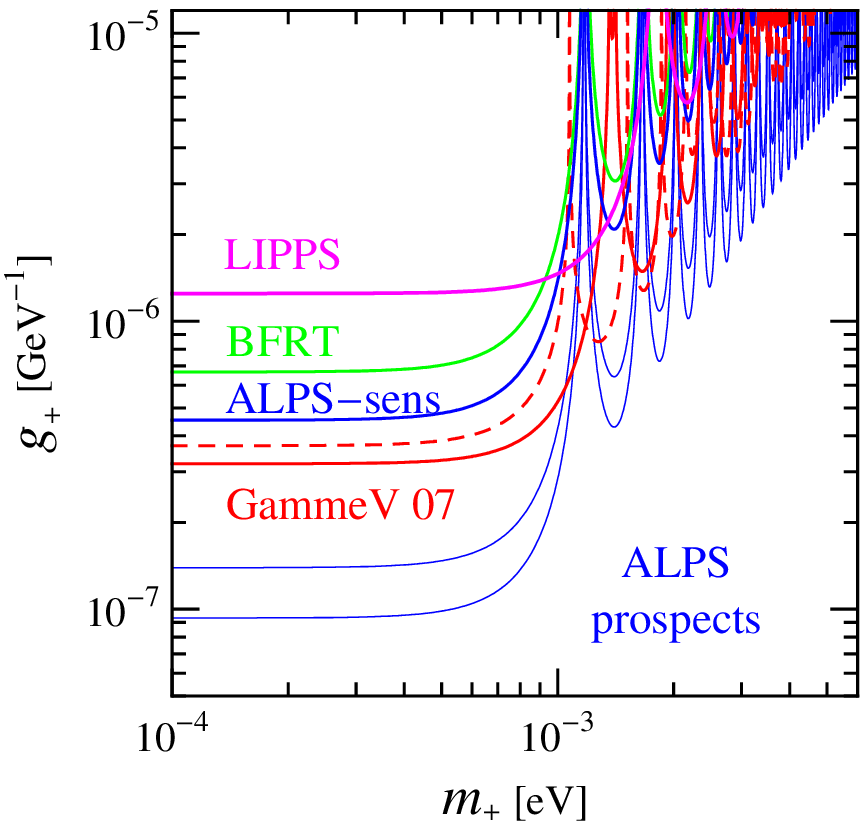}  \includegraphics[width=0.5\textwidth]{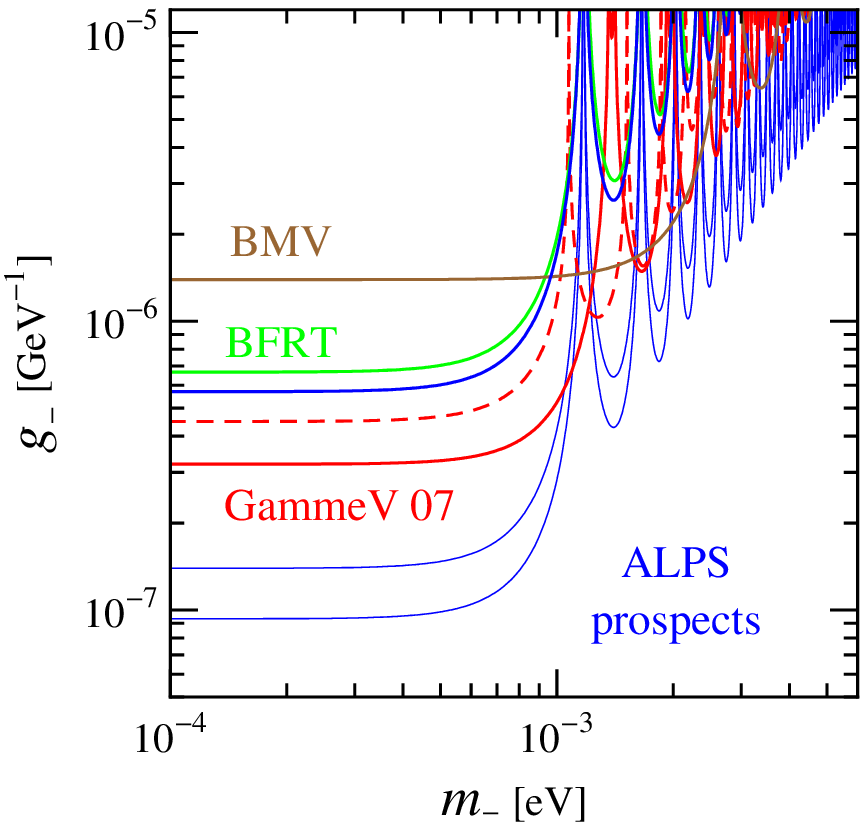} }
\centerline{ \includegraphics[width=0.6\textwidth]{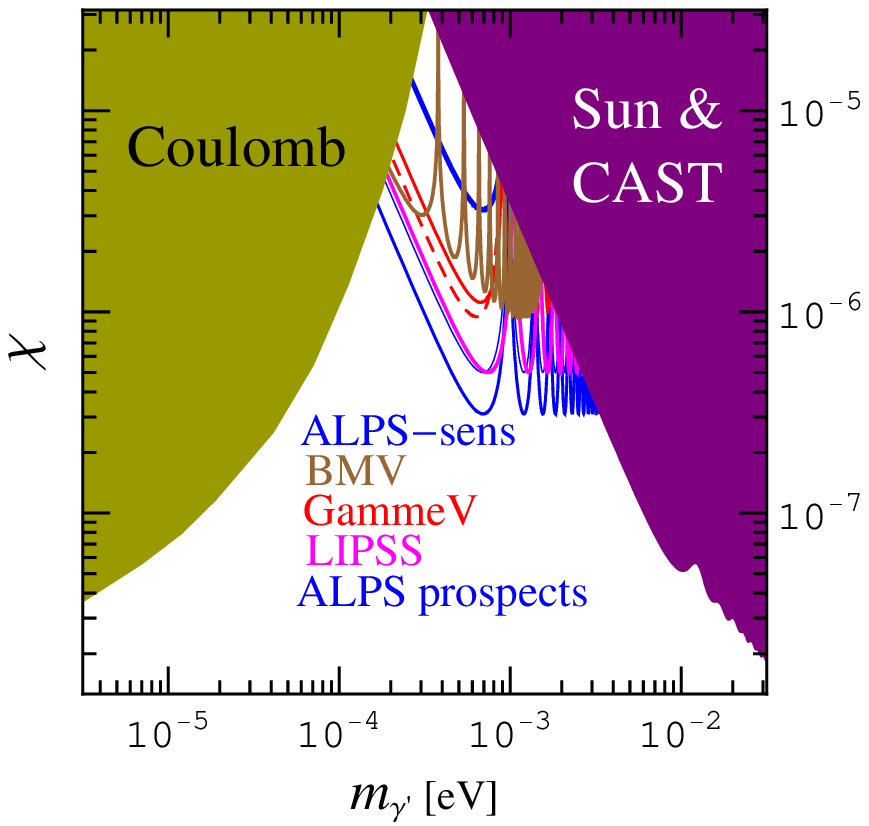}}
\caption{Comparison of sensitivities of ALPS and other axion-like-particle searches for scalar ALPs,  pseudo scalar ALPs and   
``hidden sector photons"~\cite{Amsler:2008zz}.}\label{Fig:Sens}
\end{figure}

Exploiting the two major improvements of the setup which are under preparation, the resonant cavity and a high performance camera, increases the sensitivity of the setup by nearly one order of magnitude. 
The lower curves in Fig.~\ref{Fig:Sens} show the expected ALPS sensitivity with 300~W of green laser light and this together with the additional improved detector.
We expect to have the improved setup by the end of the year in operation, allowing us to explore 
yet untouched areas in the parameter space of the low energy frontier. 

\section*{Acknowledgments}
The ALPS collaboration gratefully acknowledges the large interest and great support of many DESY groups for our activities. 
We thank especially the technical staff of the MKS group for the operation of the magnet and the Helmholtz Association for their financial support.  
 

\begin{footnotesize}

\providecommand{\href}[2]{#2}\begingroup\raggedright\endgroup


\end{footnotesize}


\end{document}